\documentclass[twocolumn,showpacs,preprintnumbers,amsmath,amssymb,prb]{revtex4-1}
\usepackage{graphicx}
\usepackage{dcolumn}
\usepackage{bm}
\usepackage{color}
\usepackage{physics}
\usepackage{multirow}

\renewcommand{\v}{{\bf v}}

\renewcommand{\k}{{\bm{k}}}
\newcommand{\q}{{\bm{q}}}
\renewcommand{\r}{{\bm{r}}}

\newcommand{\rr}{{\bm{r}}}

\usepackage[normalem]{ulem}

\def\lsim{\lower.35em\hbox{$\stackrel{\textstyle<}{\textstyle\sim}$}}
\def\gsim{\lower.35em\hbox{$\stackrel{\textstyle>}{\textstyle\sim}$}}

\begin{document}

\title{Universal mechanism of Ising superconductivity in twisted bilayer, trilayer and quadrilayer graphene}

\author{J. Gonz\'alez$^{1}$ and T. Stauber$^{2}$}

\affiliation{
$^{1}$ Instituto de Estructura de la Materia, CSIC, E-28006 Madrid, Spain\\
$^{2}$ Instituto de Ciencia de Materiales de Madrid, CSIC, E-28049 Madrid, Spain  }
\date{\today}

\begin{abstract}
We show that the superconducivity in twisted graphene multilayers originates from a common feature, which is the strong valley symmetry breaking characteristic of these moir\'e systems at the magic angle. This leads to a breakdown of the rotational symmetry of the flat moir\'e bands down to $C_3$, and to ground states in which the time-reversal symmetry is broken for a given spin projection. However, this symmetry can be recovered upon exchange of spin-up and spin-down electrons, as we illustrate by means of a self-consistent microscopic Hartree-Fock resolution where the states for the two spin projections acquire opposite sign of the valley polarization. There is then a spin-valley locking by which the Fermi lines for the two spin projections are different and related by inversion symmetry. This effect represents a large renormalization of the bare spin-orbit coupling of the graphene multilayers, lending protection to the superconductivity against in-plane magnetic fields. In the twisted bilayer as well as in  trilayer and quadrilayer graphene, the pairing glue is shown to be given by the nesting between parallel segments of the Fermi lines which arise from the breakdown of symmetry down to $C_3$. This leads to a strong Kohn-Luttinger pairing instability, which is relevant until the Fermi line recovers gradually a more isotropic shape towards the bottom of the second valence band, explaining why the superconductivity fades away beyond three-hole doping of the moir\'e unit cell.      
\end{abstract}
\maketitle

{\it Introduction.---} 
The origin of superconductivity (SC) in twisted multilayer graphene samples\cite{Cao18a,Cao18b,Yankowitz19,Kerelsky19} is still highly debated and poses one of the major theoretical challenges of condensed matter physics.\cite{Codecido19,Shen19,Lu19,Chen19,Xu18,Volovik18,Yuan18,Po18,Roy18,Guo18,Dodaro18,Liu18,Slagle18,Peltonen18,Kennes18,Koshino18,Kang18,Isobe18,You18,Wu18b,Zhang18,Ochi18,Thomson18,Carr18,Guinea18,Zou18,Laksono18,Gonzalez19,Kang19,Gonzalez19b,Carr20,Lopez-Bezanilla20,Cao21,Fischer21,LiuXiaoxue22,Christos22,Scammell22} Whereas initially also electron-phonon pairing was intensively discussed, it has now become an experimental evidence that the electron-electron interaction is responsible for the unconventional low-temperature behavior of twisted bilayer graphene (TBG) at the magic angle.

In this paper, we establish a common ground to understand the interaction effects in TBG, twisted trilayer graphene (TTG), and twisted quadrilayer graphene (TQG). We elucidate that the SC originates in all of them from a common feature, which is the strong valley symmetry breaking (VSB) characteristic of these moir\'e systems at the magic angle. Such a strong symmetry breaking pattern has profound consequences and, in TBG as well as for TTG and TQG, it leads to the breakdown of the $C_6$ symmetry of the quasi-flat moir\'e bands down to $C_3$, which is the natural symmetry group operating in a single valley. 

The breakdown of inversion symmetry gives rise to a strong distortion of the quasi-flat valence bands (VBs) in the Brillouin zone, with a strong modulation of the Coulomb interaction along the anisotropic Fermi line which reflects in the development of some negative (attractive) couplings in the decomposition over the different harmonics. This type of Kohn-Luttinger (KL) pairing instability is relevant until the Fermi line recovers a more isotropic shape towards the bottom of the second VB, which explains the reduction of the critical temperature of SC beyond three-hole doping per moir\'e unit cell.

The strong VSB can be confirmed from the experimental observations in TBG and TTG. It offers an explanation for the reset of the Hall density around  filling fraction $\nu = -2$, which otherwise would be an ordinary filling level in the middle of the lowest-energy VBs. At $\nu = -2$, the strong VSB splits the two valleys, placing the Fermi level at the vertices of the Dirac cones in the energetically lower valley. The Dirac nodes are destabilized by a sufficiently strong Coulomb interaction, opening a gap through a mechanism of dynamical symmetry breaking. This subsequent pattern turns out to be time-reversal symmetry breaking (TRSB) driving into an insulator phase in the case of TTG, thus explaining the experimental observation of a Chern number $C = -2$ in the trilayer at $\nu = -2$. However, the subsequent metallic or insulating character (after TRSB) is less robust in TBG and it may fluctuate depending on the twist angle and the screening environment, making more variable the experimental observations.

The lack of inversion symmetry in the regime of VSB implies also that singlet Cooper pairs can only be formed if electronic states with opposite spin projection are allowed to have opposite sign of the valley polarization, so that the two spin projections are attached to opposite valleys. This type of solution for the two spin sectors arises naturally in the self-consistent resolution of the interacting system in the Hartree-Fock approximation, as we show below.

\begin{figure}[t]
\includegraphics[width=0.45\columnwidth]{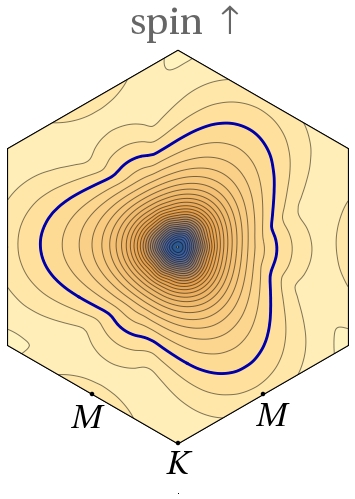}
\includegraphics[width=0.45\columnwidth]{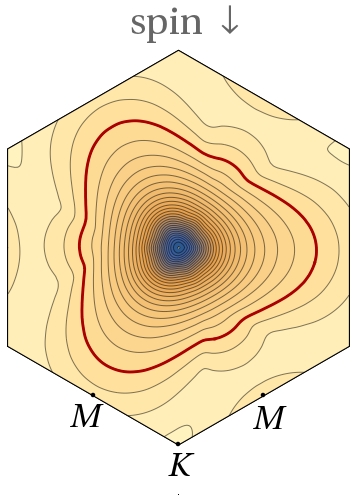}
\\
\hspace*{1.0cm} (a) \hspace{3.7cm} (b)
\caption{Energy contour maps showing the Fermi lines in the second valence band for (a) spin-up and (b) spin-down electrons in the moir\'e Brillouin zone of TBG with twist angle $\theta \approx 1.08^\circ$, for interaction strength $e^2/4\pi \epsilon = 0.1$ eV$\times a$ (the same units as in Fig. \ref{two}) and filling fraction of 2.4 holes per moir\'e unit cell. Contiguous contour lines differ by a constant step of 0.2 meV, from lower energies in blue to higher energies in light color.}
\label{one}
\end{figure}

VSB leads then to spin-valley locking, where the Fermi lines for the two spin projections are different and related by inversion symmetry, as shown in Fig. \ref{one}. This leads to a particular realization of the so-called Ising SC, as the spin-valley locking represents a large renormalization of the small spin-orbit coupling of the graphene multilayers, lending protection to the SC against in-plane magnetic fields. This may explain the large in-plane fields required to destroy the SC and the Pauli limit violation observed in TTG, and which would be absent in TBG due to the large orbital Zeeman coupling characteristic of the bilayer system.

{\it Symmetry breaking in Hartree-Fock approximation.---}
We carry out a real space description of the twisted multilayers, taking into account all the carbon atoms in the supercell of the moir\'e systems. In this respect, we are going to consider some representatives from the sequence of commensurate superlattices with twist angle $\theta_i=\arccos ((3i^2+3i+1/2)/(3i^2+3i+1))$. In the case of TBG we will focus on the system with $i=30$, with a twist angle $\theta = 1.08^\circ$ which matches well the magic-angle condition. We will recall the results obtained for TTG in Ref. \onlinecite{Gonzalez22}, which considered the configuration with alternating twist angle between layers with $\theta = 1.61^\circ$, very close to the magic-angle condition for this symmetric TTG. In the case of TQG, we will assemble two twisted bilayers with $i=18$ and alternating twist angle $\theta = 1.79^\circ$, which place the system very close to the flat-band condition. 

The starting point in the real space approach is the tight-binding approximation for the noninteracting system, in which the Hamiltonian $H_0$ is taken with a typical parametrization of the exponential decay of hopping matrix elements, as given in Ref. \onlinecite{Moon13}. To this, we add the interacting Hamiltonian $H_{\rm int}$ accounting for the Coulomb interaction, which is expressed in terms of creation (annihilation) operators $a_{i\sigma}^{\dagger}$ ($a_{i\sigma}$) for electrons at each carbon site $i$ with spin $\sigma$ as
\begin{align}  
H_{\rm int} = \frac{1}{2} \sum_{i,j,\sigma,\sigma'} a_{i\sigma}^{\dagger}a_{i\sigma} \: v_{\sigma \sigma'} (\r_i-\r_j) \: a_{j\sigma'}^{\dagger}a_{j\sigma'}\;,
\end{align}    
For $\r_i \neq \r_j$, we take $v_{\sigma \sigma'} (\r_i-\r_j)=v(\r_i-\r_j)$, $v$ being the extended Coulomb potential. This is supposed to be screened by metallic gates above and below the graphene compound, which introduces an exponential decay with screening length $\xi$. Moreover, the strength of the potential is also reduced by a dielectric constant $\epsilon $, accounting mainly for internal screening from electron-hole excitations. For $\r_i = \r_j$, we have the Hubbard interaction $v_{\sigma \sigma'} = U \delta_{\sigma,-\sigma'}$. The precise value of this coupling is not very important, as long as it is nonvanishing, but it plays an essential role to constrain the relative orientation of the spin projections in the two valleys of the multilayers.

We adopt a Hartree-Fock (HF) approach to the many-body problem, which implies the self-consistent resolution of the Dyson equation involving the interacting (noninteracting) electron propagator $G$ ($G_0$) and the self-energy $\Sigma $
\begin{align}
G^{-1} = G_0^{-1} - \Sigma 
\label{sd}
\end{align}
The HF approximation proceeds by assuming that the static limit of the propagator $G$ admits a representation similar to that of $G_0$, but with a set of eigenvalues $\varepsilon_{a\sigma}$ and eigenvectors $\phi_{a\sigma} (\r_i)$ modified by the interaction:
\begin{align}
\left(  G  \right)_{i\sigma,j\sigma} = -\sum_a \frac{1}{\varepsilon_{a\sigma}}  \phi_{a\sigma} (\r_i)  \phi_{a\sigma} (\r_j)^*\;.
\label{hf}
\end{align}
The resolution of (\ref{sd}) is then feasible since the self-energy is given in terms of the eigenvectors by
\begin{align}
\left( \Sigma  \right)_{i\sigma,j\sigma}  = &  \; \mathbb{I}_{ij} \:  \sideset{}{'}\sum_a  \sum_{l, \sigma' } v_{\sigma \sigma'} (\r_i-\r_l)   \left|\phi_{a\sigma'} (\r_l)\right|^2      \notag    \\ 
&  - v_{\sigma \sigma} (\r_i-\r_j)  \sideset{}{'}\sum_a \phi_{a\sigma} (\r_i)  \phi_{a\sigma} (\r_j)^*\;,
\label{selfe}
\end{align}
where the prime denotes that the sum is only over the occupied levels \cite{Fetter71}.  

The HF approach is well-suited to our discussion, since the Fock potential is the essential contribution triggering the different symmetry-breaking order parameters. These can be written from the matrix elements 
\begin{align}
h_{ij}^{(\sigma )} =  \sideset{}{'}\sum_a \phi_{a\sigma} (\rr_i) \phi_{a\sigma} (\rr_j)^*\;.
\end{align}
Thus, the order parameters for TRSB are given by
\begin{align}
P_{\pm}^{(\sigma )} = {\rm Im} \left( \sum_{i \in A}  \left(   h_{i_1 i_2}^{(\sigma )} h_{i_2 i_3}^{(\sigma )} h_{i_3 i_1}^{(\sigma )}   \right)^{\frac{1}{3}}
\pm \sum_{i \in B}  \left(   h_{i_1 i_2}^{(\sigma )} h_{i_2 i_3}^{(\sigma )} h_{i_3 i_1}^{(\sigma )}   \right)^{\frac{1}{3}}  \right)
\label{ssigma}
\end{align}
where the sums run over the loops made of three nearest neighbors $i_1, i_2$ and $i_3$ of each atom $i$ in graphene sublattices $A$ and $B$. A nonvanishing $P_+$ corresponds to the usual Chern insulator phase with Haldane mass, while $P_- \neq 0$ signals instead an uneven rigid shift in the energies of the two valleys of the moir\'e system. Moreover, the other pattern likely to be found corresponds to chiral symmetry breaking, measured by the order parameter
\begin{align}
C^{(\sigma )} = & \sum_{i \in A} h_{ii}^{(\sigma )}  - \sum_{i \in B} h_{ii}^{(\sigma )} 
\label{chiral}
\end{align}

\begin{figure}[h]
(a)$ $\hspace{8cm}$ $\\
\includegraphics[width=0.75\columnwidth]{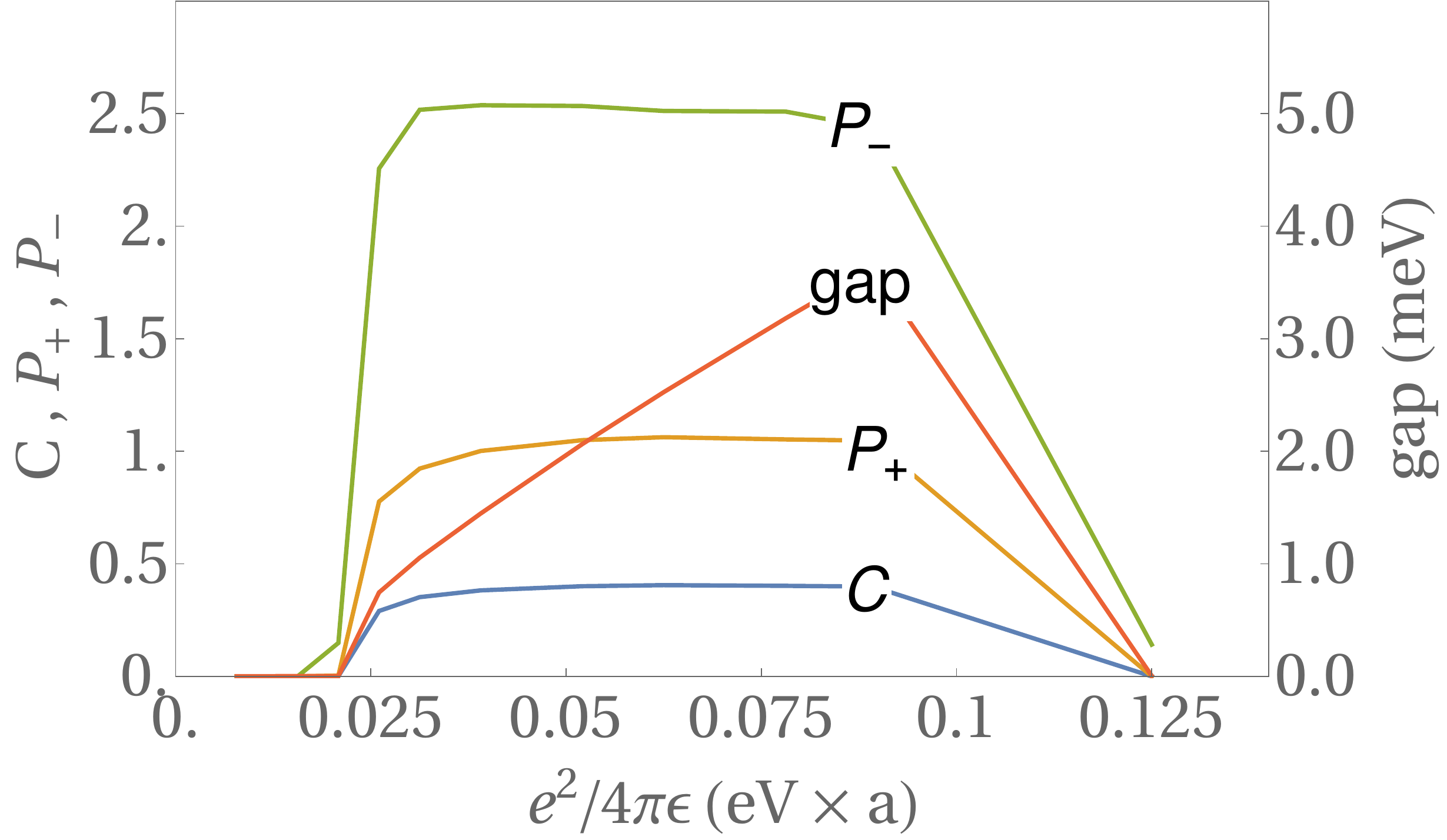}\\
(b)$ $\hspace{8cm}$ $\\
\includegraphics[width=0.75\columnwidth]{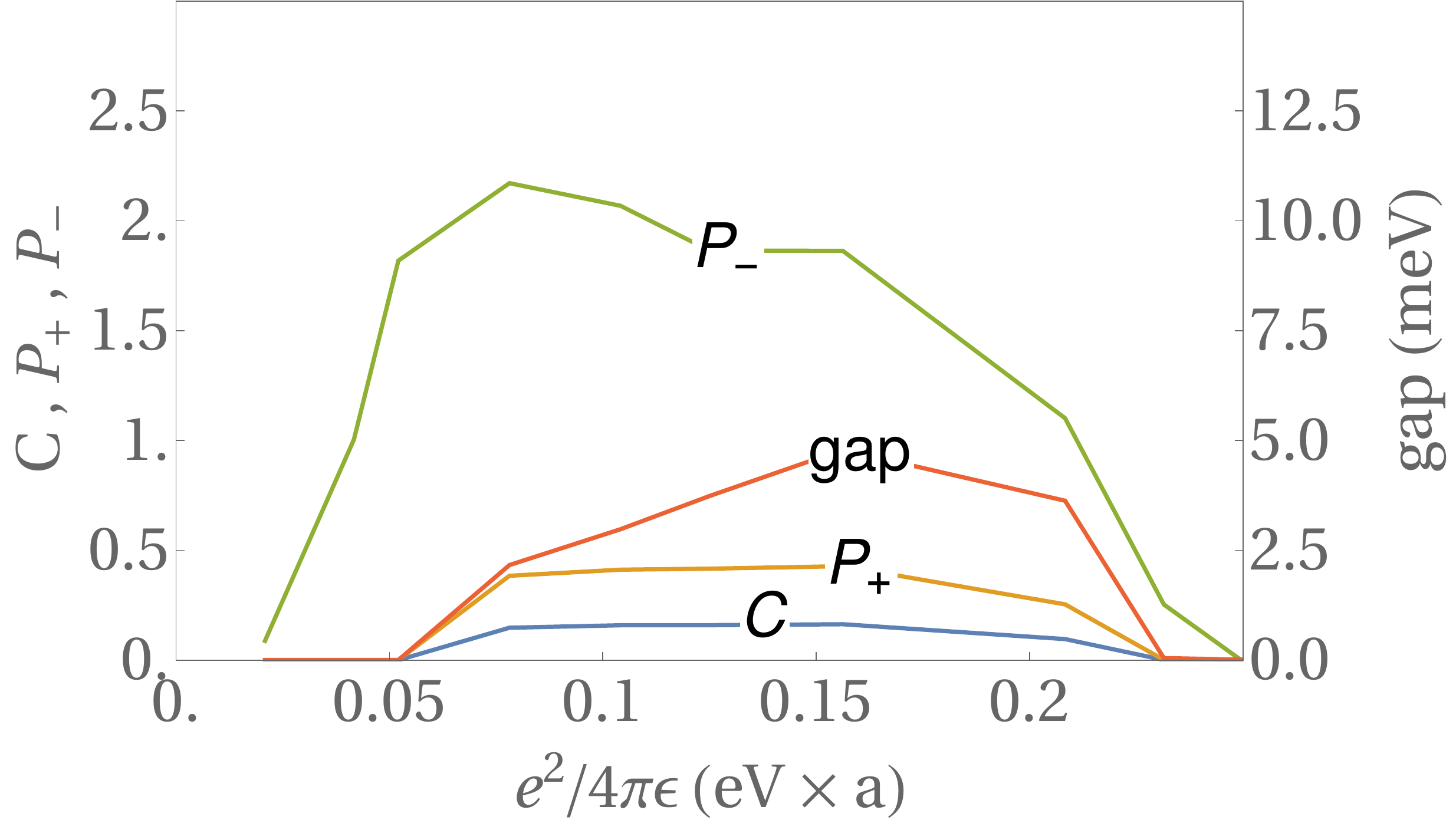}
\caption{Phase diagrams showing the order parameters of symmetry breaking at a filling fraction of 2 holes per moir\'e unit cell in (a) TBG at twist angle $\theta \approx 1.08^\circ$ and (b) TQG at $\theta \approx 1.79^\circ$, obtained by means of a self-consistent Hartree-Fock approximation. In the two cases the screening length of the long-range Coulomb potential is $\xi = 10$ nm. The interaction strength is measured in units of eV times the C-C distance $a$.}
\label{two}
\end{figure}

The resolution of Eq. (\ref{sd}) allows us to map the different symmetry breaking order parameters as the strength of the Coulomb interaction evolves with $\epsilon $, from weak to strong coupling. The different phases obtained at filling fraction $\nu = -2$ for TBG and TQG at their respective magic angles are shown in Fig. \ref{two}. The phase diagram for TTG with $\theta = 1.61^\circ$ at the same hole doping can be found in Ref. \onlinecite{Gonzalez22}.

It is remarkable that the picture of symmetry breaking patterns is qualitatively the same in the three multilayers. The critical point for the opening of a gap at the Dirac nodes is shifted towards weaker coupling in Fig. \ref{two}(a), but this is due to the fact that TBG at $\theta = 1.08^\circ$ starts from a more critical condition, with a width of the noninteracting quasi-flat bands smaller than in the other two cases. On the other hand, we observe that another transition takes place to the right of the phase diagrams. The order parameter of VSB vanishes at a certain interaction strength, which is the signal that VSB is giving way to a different phase characterized by intervalley coherence in the strong coupling limit. However, in the absence of valley polarization, this new pattern cannot account for the jumps observed in the Hall density of TBG and TTG at $\nu = -2$, so we will not consider such a strong coupling phase in the subsequent discussion.

The common structure of the phases in the three multilayers is indeed ubiquitous in the moir\'e systems, as it is also found in TBG at larger twist angles where the magic-angle condition is tuned by applying hydrostatic pressure. These systems with smaller supercells afford an exact HF treatment, including all the valence bands. In the real systems with 5000-10000 atoms in the moir\'e, we only include up to 128 bands around the neutrality point in the self-consistent resolution, due to technical reasons. Nevertheless, the exact resolutions of the smaller systems show a similar structure of the phases as in Figs. \ref{two}(a)-(b) which lends further support to the results obtained here.

\begin{figure}[h]
\includegraphics[width=0.45\columnwidth]{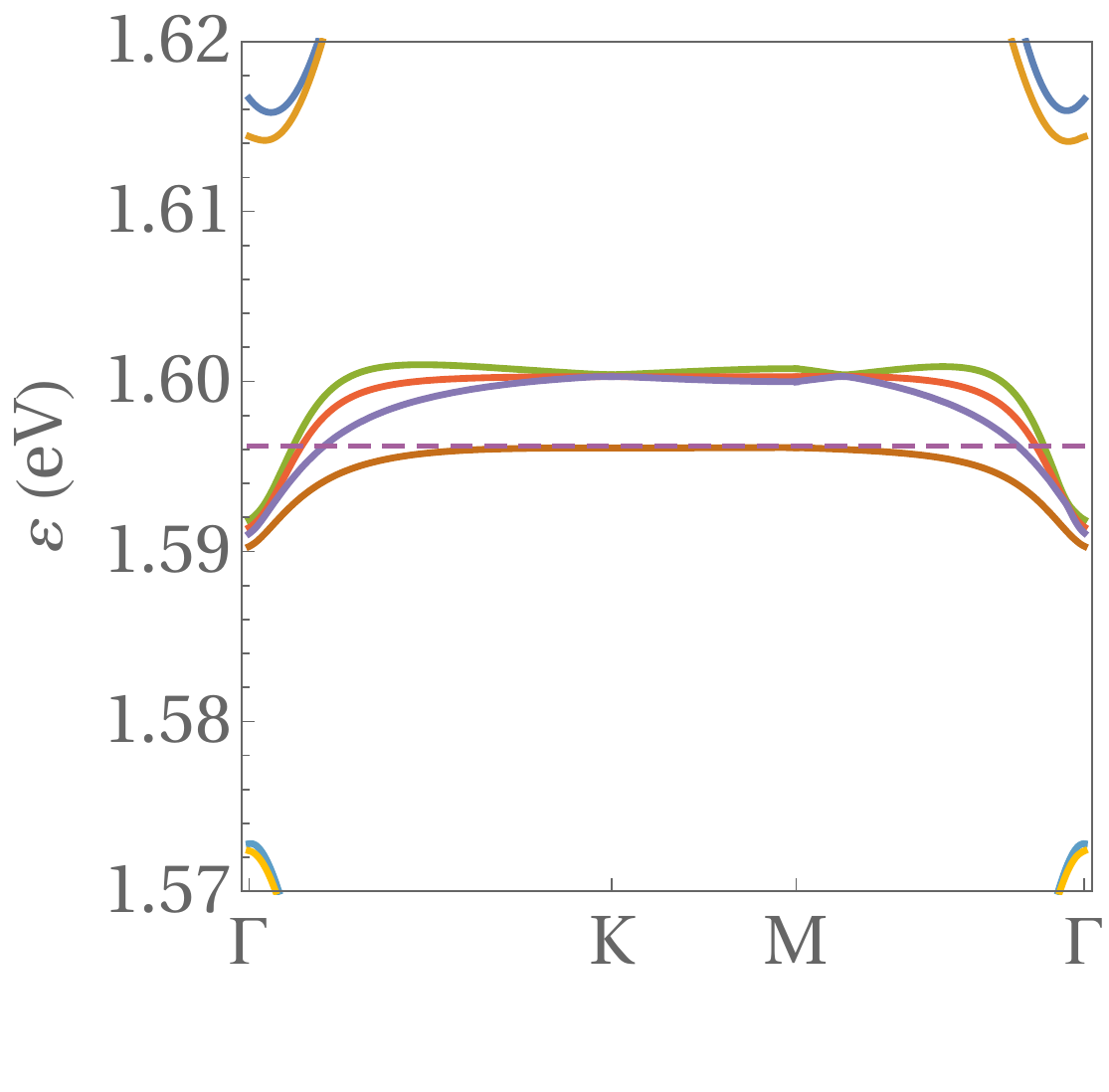}
\includegraphics[width=0.45\columnwidth]{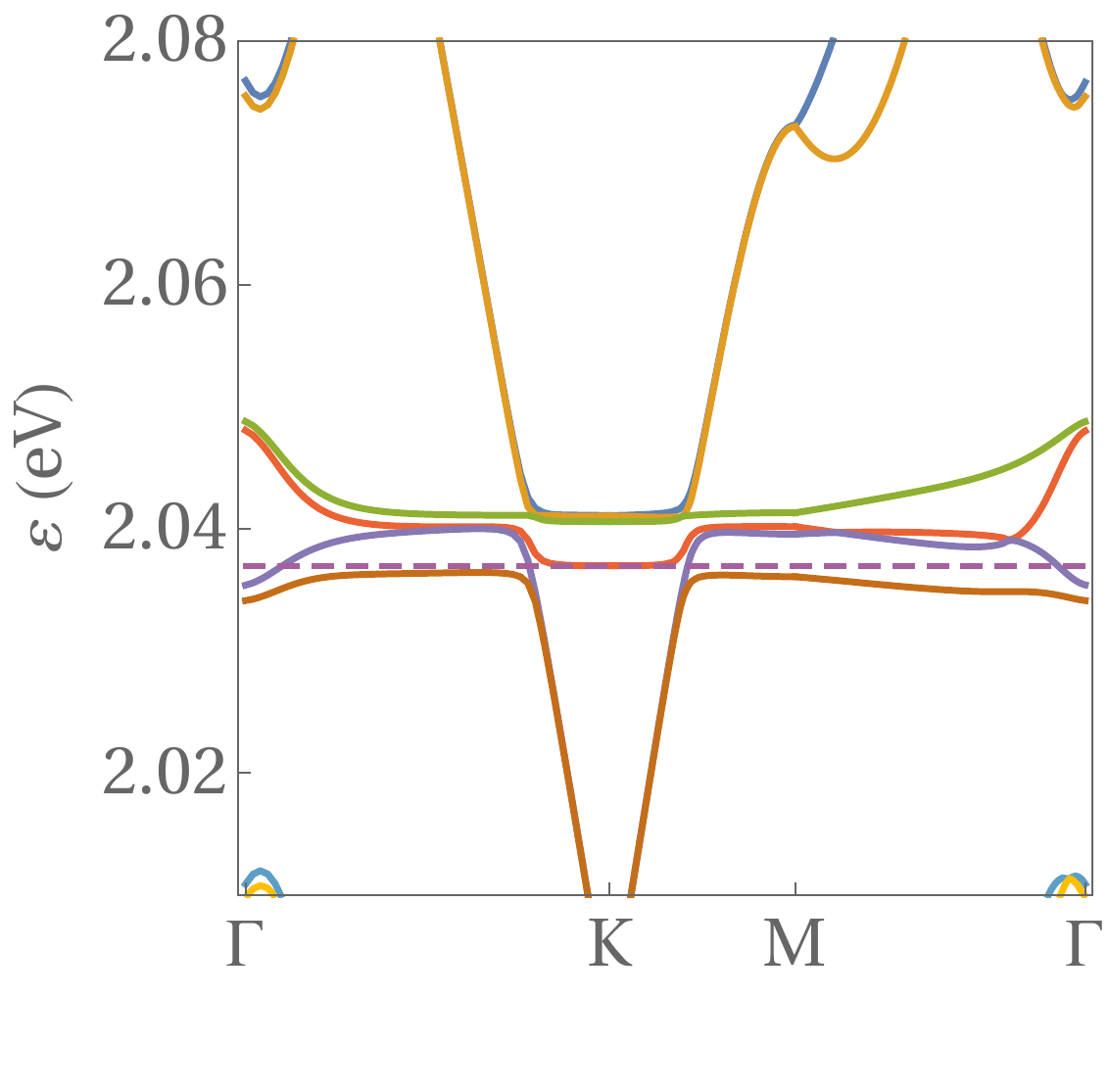}
\\
\hspace*{1.0cm} (a) \hspace{3.7cm} (b)
\caption{Lowest valence and conduction bands of (a) TBG at twist angle $\theta \approx 1.08^\circ$ and (b) TQG at twist angle $\theta \approx 1.79^\circ$, computed in a self-consistent Hartree-Fock approximation for respective interaction strengths $e^2/4\pi \epsilon = 0.1, 0.125$ eV$\times a$, at filling fraction of 2 holes per moir\'e unit cell (the dashed line stands for the Fermi level).}
\label{three}
\end{figure}

We remark that the gap plotted in Figs. \ref{two}(a)-(b) corresponds to the separation between valence and conduction bands at the $K$ point in the lower valley, in the presence of valley polarization. We consider that TBG, as well as TTG and TQG, must be placed in the phase with VSB, since the splitting of the conduction and the valence band in the lower valley is the feature that explains the jump in the Hall density observed experimentally at $\nu = -2$. In the phase with VSB, TBG and TQG show indeed at $\nu = -2$ a splitting of the bands at the Fermi level, as can be seen in Fig. \ref{three}. It has been shown in Ref. \onlinecite{Gonzalez22} that a similar effect takes place in the case of TTG, and that the splitting of the valence and the conduction band reproduces the reset of the Hall density found experimentally near the integer filling.      

Let us finally comment on the behavior of the band-structure of TBG close to the $\Gamma$ point. As can be seen in Fig. \ref{three}(a), the eigenenergies are shifted to lower energies in contrary to the expectation that the Coulomb energy should push them upward. The reason is that the Hartree contribution is not fully taken into account as only 128 bands are considered. Nevertheless, it is the exchange energy that is responsible for the symmetry breaking which is treated adequately.

\begin{figure}[t]
\includegraphics[width=0.45\columnwidth]{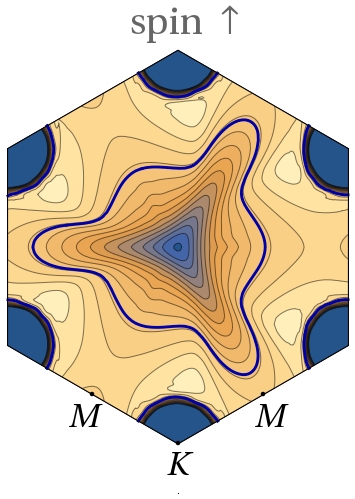}
\includegraphics[width=0.45\columnwidth]{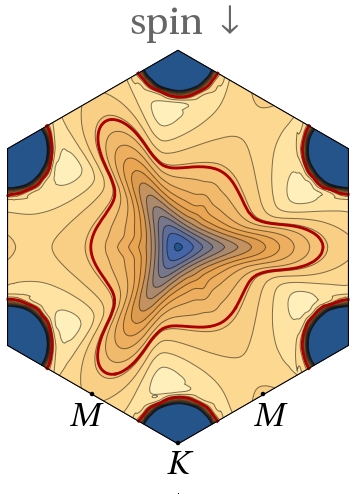}
\\
\hspace*{1.0cm} (a) \hspace{3.7cm} (b)
\caption{Energy contour maps showing the Fermi lines in the second valence band for (a) spin-up and (b) spin-down electrons in the moir\'e Brillouin zone of TQG with twist angle $\theta \approx 1.79^\circ$, for interaction strength $e^2/4\pi \epsilon = 0.125$ eV$\times a$ (the same units as in Fig. \ref{two}) and filling fraction of 2.8 holes per moir\'e unit cell. Contiguous contour lines differ by a constant step of 0.2 meV, from lower energies in blue to higher energies in light color.}
\label{four}
\end{figure}

{\it Ising superconductivity.---}
The strong VSB leads to ground states in which the time-reversal symmetry is broken for a given spin projection, but where this symmetry is recovered upon exchange of the two spin projections. This is indeed realized in the HF approximation as this approach allows for solutions where the states for the two spin sectors have opposite sign of the VSB order parameter. The bands are then symmetric under the transformation $\k \rightarrow -\k $ and concurrent exchange of the spin projections, as shown in Figs. \ref{one} and \ref{four}. We observe in the figures that the two spin sectors have different Fermi lines, related by inversion symmetry in momentum space.

The main consequence of the spin-valley locking is that the electrons with opposite momenta of a Cooper pair are forced to live on different Fermi lines attached to opposite valleys. In this regard, this effect represents a strong reinforcement of the small spin-orbit coupling in the graphene layers. A similar renormalization was considered in Ref. \onlinecite{Kane05}, taking into account diagrammatic contributions which coincide with those encoded in the Fock part of our approximation. This implies that the SC in the twisted multilayers must be of Ising type, as the spin-valley locking favors a preferred orientation of the spins of a Cooper pair, normal to the graphene planes.

What remains to be discussed is the pairing instability in the system, which demands the analysis of the Cooper pair vertex $V$ for electrons with zero total momentum. This vertex can be parametrized in terms of the angles $\phi $ and $\phi'$ of the respective momenta of the spin-up incoming and outgoing electrons on each contour line of energy $\varepsilon $. The instabilities of the vertex can be obtained by solving the equation encoding the iteration of the scattering of Cooper pairs:
\begin{eqnarray}
&&V(\phi, \phi')= V_0 (\phi, \phi')-   \nonumber        \\ 
 && \frac{1}{(2\pi )^2} \int^{\Lambda_0} \frac{d \varepsilon }{\varepsilon } 
   \int_0^{2\pi } d \phi'' 
  \frac{\partial k_\perp }{\partial \varepsilon}  
      \frac{\partial k_\parallel }{\partial \phi''} 
  V_0 (\phi, \phi'')    
         V(\phi'', \phi')
\label{pp}
\end{eqnarray}
where $k_\parallel ,k_\perp $ are the longitudinal and transverse components of the momentum for each energy contour line while $V_0 (\phi, \phi') $ is the bare vertex at a high-energy energy cutoff $\Lambda_0$.

Eq. (\ref{pp}) can be simplified by differentiating with respect to the cutoff, which leads to
\begin{equation}
\varepsilon \: \frac{\partial \widehat{V}(\phi, \phi' )}{\partial \varepsilon } 
 =   \frac{1}{2\pi }  \int_0^{2\pi } d \phi''  
 \widehat{V} (\phi, \phi'' )  \widehat{V}(\phi'', \phi' )
\label{scaling}
\end{equation}
with $\widehat{V} (\phi, \phi' ) = F(\phi ) F(\phi' ) V (\phi, \phi' )$ and $F(\phi ) = \sqrt{ (\partial k_\perp / \partial \varepsilon )
  (\partial k_\parallel / \partial \phi )/2\pi  }$. Eq. (\ref{scaling}) implies that the vertex is a function of the variable $\varepsilon/\Lambda_0$. If the initial condition $V_0 (\phi, \phi')$ has a negative eigenvalue for any of its projections (harmonics) over the Fermi line, the solutions of (\ref{scaling}) will display then a divergence in the magnitude of the negative coupling as $\varepsilon \rightarrow 0$, which is the signature of the pairing instability.  

\begin{figure}[h!]
\includegraphics[width=0.2\columnwidth]{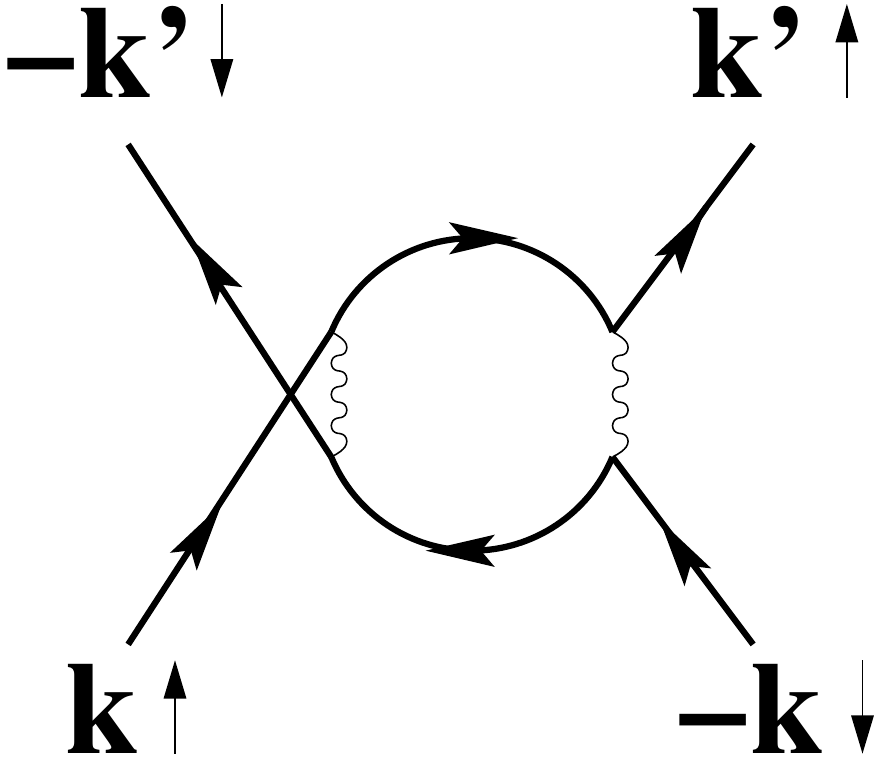}
\caption{Lowest-order diagram dressing the BCS vertex in the particle-hole channel.}
\label{five}
\end{figure}

The crucial question is to start with a sensible representation of the initial vertex $V_0 (\phi, \phi')$ at the high-energy cutoff. For this we rely on the iteration of diagrams in the particle-hole channel shown in Fig. \ref{five}, which are not included in the RPA diagrams already accounted for by the internal screening of the Coulomb potential $v$. We have therefore\begin{equation}
V_0 (\phi, \phi') =  v_{\k-\k^\prime}
    + \frac{v_{\bm{Q}}^2 \: \widetilde{\chi}_{\k+\k^\prime }}{1 - v_{\bm{Q}} \: \widetilde{\chi}_{\k+\k^\prime }}   \;,
\label{init}
\end{equation} 
where $\k, \k^\prime$ are the respective momenta for angles $\phi, \phi'$ and $\widetilde{\chi}_{\q}$ is particle-hole susceptibility appearing in Fig. \ref{five}.

The final step is to project the vertex onto the harmonics $\cos (n\phi),\sin (n\phi)$ which build up the different contributions to $\widehat{V} (\phi, \phi' )$ at the high-energy cutoff. We have carried out this operation along the Fermi lines of TBG shown in Fig. \ref{one}, at filling fraction $\nu = -2.4$. The eigenvalues for the different harmonics can be grouped according to the irreducible representations of the approximate symmetry group $C_{3v}$. The results can be seen in Table \ref{table1}.

\begin{table}[t!]
\begin{tabular}{c|c|c}
Eigenvalue $\lambda$  &   harmonics  &   Irr. Rep.\\
\hline \hline
4.25  &     $     1     $         &                  \\
\hline
2.14   &    \multirow{2}{*}{ $\{\cos (\phi),\sin (\phi)\}$ }    & \multirow{2}{*}{E}  \\
2.09   &                                  &                                      \\

\hline
$-0.43$  &    $ \{\cos (7\phi),\sin (7\phi), $   &  \multirow{2}{*}{E}  \\
$-0.40$   &    $  \cos (8\phi),\sin (8\phi)\} $    &                                         \\

\hline
$-0.36$  &     $     \cos (6\phi)     $         &       $A_1$            \\
\hline
0.33   &     $\cos (3\phi)  $       &     $A_1$               \\                
\hline
0.26   &   $ \{\cos (2\phi),\sin (2\phi), $    &  \multirow{2}{*}{E}   \\
0.21   &    $ \cos (5\phi),\sin (5\phi)\}  $   &                                               \\

\end{tabular}
\caption{Eigenvalues of the Cooper-pair vertex with largest magnitude and dominant harmonics grouped according to the irreducible representations of the approximate $C_{3v}$ symmetry, for the Fermi line of TBG shown in Fig. \ref{one}.} 
  \label{table1}
\end{table}

We find that, among the eigenvalues in Table \ref{table1}, there is a pair of negative couplings with relatively large magnitude $|\lambda| \approx 0.4$. We recall that, for a negative coupling $\lambda $ in a given representation, the solution of Eq. (\ref{scaling}) develops a divergence in that channel at the energy scale
\begin{equation}
\varepsilon_c = \Lambda_0 \: e^{-1/|\lambda |}
\end{equation}
In TBG, the magnitude of $\Lambda_0 $ is constrained by the reduced bandwidth of the second VB in Fig. \ref{one}. We can assign to $\Lambda_0 $ the value of half the bandwidth, so that $\Lambda_0 \sim 1$ meV. In this case, the small magnitude of the cutoff is compensated by the relatively large value of $|\lambda |$, which leads to a critical temperature $T_c \sim 1$ K.

\begin{table}[h!]
\begin{tabular}{c|c|c}
Eigenvalue $\lambda$  &   harmonics  &   Irr. Rep.\\
\hline \hline
3.54  &     $     1     $         &                  \\
\hline
1.40   &   $ \{\cos (\phi),\sin (\phi), $    &  \multirow{2}{*}{E}   \\
1.39   &    $ \cos (2\phi),\sin (2\phi)\}  $   &                                               \\
\hline
0.37   &     $\cos (3\phi)  $       &     $A_1$               \\     
\hline
$-0.31$  &    $ \{\cos (5\phi),\sin (5\phi), $   &  \multirow{2}{*}{E}  \\
$-0.30$   &    $  \cos (7\phi),\sin (7\phi)\} $    &                                         \\
\hline
$-0.30$  &     $     \cos (6\phi)     $         &       $A_1$            \\

\hline
0.27   &     $     \sin (6\phi)       $     &      $A_2$                                 \\

\end{tabular}
\caption{Eigenvalues of the Cooper-pair vertex with largest magnitude and dominant harmonics grouped according to the irreducible representations of the approximate $C_{3v}$ symmetry, for the Fermi line of TQG shown in Fig. \ref{four}.} 
  \label{table2}
\end{table}

The same analysis of Fourier decomposition can be done in the case of TQG, with the results shown in Table \ref{table2} for the Fermi line in Fig. \ref{four} at $\nu = -2.8$. The value of the dominant negative coupling is somewhat smaller than in TBG, $|\lambda | \approx 0.3$, but the estimate of the critical temperature gives the same order of magnitude $T_c \sim 1$ K. In both cases, we observe that the main limitation in $T_c$ may come from the reduced space for the scattering of the Cooper pairs, which is inherent to the small bandwidth of the second VB.

{\it Conclusion.---}
In both TBG and TQG, the large values of the dominant attractive couplings are a consequence of the approximate nesting between parallel segments in the Fermi lines shown in Figs. \ref{one} and \ref{four}. In this respect, the Kohn-Luttinger mechanism operates in rather similar way as in TTG, where there is approximate nesting between the sides of the triangular patches of the Fermi line (as shown in Ref. \onlinecite{Gonzalez22}). In the three cases, the nesting features disappear as the hole doping increases and the Fermi lines recover a more isotropic shape, explaining why the Kohn-Luttinger instability fades away before reaching the bottom of the band.

We have seen that the Fermi lines for the two spin projections are different and related by inversion symmetry in the twisted multilayers. This leads to a particular realization of the so-called Ising SC, as the spin-valley locking represents a large renormalization of the small spin-orbit coupling of the graphene multilayers, lending protection to the SC against in-plane magnetic fields. This may explain the large in-plane fields required to destroy the SC and the Pauli limit violation observed in TTG, and which would be absent in TBG due to the large orbital Zeeman coupling characteristic of the bilayer system.

{\it Acknowledgements.}
This work was supported by the project PID2020-113164GB-I00 financed by MCIN/ AEI/10.13039/501100011033. The access to computational resources of CESGA (Centro de Supercomputaci\'on de Galicia) is also gratefully acknowledged.

\end{document}